\documentstyle[12pt]{article} 
\author{L.~Didukh, Yu.~Skorenkyy\\ 
{\small \it Ternopil State Technical University, Department of Physics}\\ 
{\small \it 56 Rus'ka Str., Ternopil UA--282001, Ukraine}\\
{\small \it E-mail: didukh@tu.edu.te.ua}} 
\date{}
\title{Electron correlations in narrow energy bands:
ground state energy and metal-insulator transition}

\begin{document}
\maketitle

\begin{abstract}
The electron correlations in narrow energy bands are examined 
in framework of the Hubbard model. The single-particle Green function
and energy spectrum are obtained in paramagnetic state at half-filling
by means of new two-pole approximation.
In the ground state analitical expressions for the energy gap, polar states 
concentration and energy of the system are found. Metal-insulator transitions
in the model at change of bandwidth or temperature are 
investigated. The obtained results are used for interpretation of some 
experimental data in narrow-band materials.
\end{abstract}
\section{Introduction}

Among the observed in narrow-bands materials metal-insulator transitions~(MIT)
the significant interest is attracted to the transitions from 
paramagnetic metal state to paramagnetic insulator state at increase of
temperature which exhibit the systems NiS$_{2-x}$Se$_x$~\cite{mott}-\cite{spa}, 
(V$_{1-x}$Cr$_x$)$_2$O$_3$~\cite{mott,edra} and 
Y$_{1-x}$Ca$_x$TiO$_3$~\cite{tag,tok}; in these systems the 
paramagnetic insulator -  paramagnetic metal transitions under external 
pressure are observed also. There are reasons to believe that noted transitions
are the consequences of electron-electron interactions and can be described 
within the framework of Hubbard model~\cite{hubb1}.

Hubbard model is the simplest model describing MIT in 
materials with narrow energy
bands. This model describes a single 
non-degenerate band of electrons with the local Coulomb interaction. The 
model Hamiltonian contains two energy parameters: 
the hopping integral of an electron from one site to another
and the intraatomic Coulomb repulsion of 
two electrons of the opposite spins. This model is used intensively (for 
recent reviews see Refs.~\cite{mott,iz}-\cite{gebh}) in order to describe 
the peculiarities of physical properties of narrow-band materials;
in this connection two-pole approaches are attractive.
The two-pole approaches in the Hubbard model and the Hubbard bands 
conception (being the consequence of two-pole approximation) have been 
useful for understanding of the peculiarities of electric and magnetic
properties of narrow-band materials~\cite{mott,gebh}. However 
within the framework of two-pole approaches there are series of issues,
in particular the problem of metal-insulator transition 
description~\cite{gebh,iked}-\cite{izyu}.

In the present paper recently proposed two-pole approximation~\cite{dcm}
is used to study effects of electron correlations in the Hubbard model.
The single particle Green function
and energy spectrum are obtained. In the ground state analitical dependences
of energy gap, polar states (doublons or holes) concentration and energy of 
the system on model parameters are found. Dependences for energy gap, 
polar states concentration on temperature are calculated. The obtained 
results are compared with corresponding results of other approximations
and are used for the interpretation of some experimental data.
In particular, the observable transitions from an insulating state 
to a metallic one at increase of bandwidth, and  from a metallic state 
to an insulating one at increasing temperature are explained.

\section{Single-particle Green function and energy spectrum}

The Hubbard Hamiltonian~[1] in terms of transition-operators 
of $i$-site from state $|l\rangle$ to state $|k\rangle$ 
$X_i^{kl}$~\cite{hubb}  is written as
\begin{eqnarray}
&&H=H_0+H_1+H'_1;\label{ham}\\
&&H_0=-\mu \sum_{i\sigma} \left(X_i^{\sigma}+X_i^2\right)+
U\sum_{i}X_i^2,\\
&&H_1=\sum_{ij\sigma,i\neq j}t_{ij}\left(X_i^{\sigma 0}X_j^{0\sigma}
+X_i^{2\bar{\sigma}}X_j^{\bar{\sigma} 2}\right),\\
&&H'_1=\sum_{ij,i\neq j}t_{ij}\left(X_i^{\bar{\sigma}0}X_j^{\sigma 2}
-X_i^{\sigma 0}X_j^{\bar{\sigma} 2}+h.c.\right),
\end{eqnarray}
where $\mu$ is the chemical potential, $U$ is the intra-atomic Coulomb repulsion,
$t_{ij}$ is the nearest-neighbor hopping integral, $X_i^k$ is the operator 
of number 
of $|k\rangle$-states on $i$-site; $\sigma$ denotes spin of an electron 
($\sigma=\downarrow,\uparrow$) and $\bar{\sigma}$ denotes the projection 
of electron spin opposite to $\sigma$; 
$H_{0}$ describes system in the atomic limit,
$H_{1}$ describes electron hoppings between single occupied sites and empty 
sites (holes) (the first sum in $H_1$ -- processes forming ``h-band'')
and electron hoppings between doubly occupied sites (doublons) and 
single occupied sites (the second sum in $H_1$ -- processes forming ``d-band'').
$H'_{1}$ describes ``hybridization'' between the ``h-band'' and ``d-band''
(the processes of pair creation and annihilation of holes and doublons).

The single-particle Green function is written in $X_i^{kl}$-operators as
\begin{eqnarray} \label{Grf}
\langle\langle a_{p\uparrow}|a^{+}_{s\uparrow}\rangle\rangle=
\langle\langle X_p^{\downarrow 2}|X_s^{2\downarrow}\rangle\rangle
- \langle\langle X_p^{0\uparrow}|X_s^{2\downarrow}\rangle\rangle 
- \langle\langle X_p^{\downarrow 2}|X_s^{\uparrow 0}\rangle\rangle 
+\langle\langle X_p^{0\uparrow}|X_s^{\uparrow 0}\rangle\rangle.  
\end{eqnarray}

The functions 
$\langle\langle X_p^{\downarrow 2}|X_s^{2\downarrow}\rangle\rangle$ and
$\langle\langle X_p^{0\uparrow}|X_s^{2\downarrow}\rangle\rangle$
satisfy the  equations
\begin{eqnarray}\label{eqs}
(E+\mu-U)\langle\langle X_p^{\downarrow 2}|X_s^{2\downarrow}\rangle\rangle&=&
{\delta_{ps}\over 2\pi}\langle X_p^{\downarrow}+X_p^{2}\rangle+ 
\langle\langle{\left[X_p^{\downarrow 2}, H_1\right]}_{-}
|X_{s}^{2\downarrow}\rangle\rangle\nonumber\\
&+&\langle\langle{\left[X_p^{\downarrow 2}, H'_1\right]}_{-}
|X_{s}^{2\downarrow}\rangle\rangle,\\
(E+\mu)\langle\langle X_p^{0 \uparrow}|X_s^{2\downarrow}\rangle\rangle&=&
\langle\langle{\left[X_p^{0 \uparrow}, H_1\right]}_{-}
|X_{s}^{2\downarrow}\rangle\rangle
+\langle\langle{\left[X_p^{0 \uparrow}, H'_1\right]}_{-}
|X_{s}^{2\downarrow}\rangle\rangle,\nonumber
\end{eqnarray} 
with ${[A, B]}_{-}=AB-BA$. To obtain the closed system of equations we 
apply new two-pole approximation, proposed in work~\cite{dcm}.
Suppose in Eq.~(\ref{eqs}) that 
\begin{eqnarray}
{\left[X_p^{0\uparrow}, H_1\right]}_{-}=\sum_{j}\epsilon(pj)X_j^{0\uparrow},
\nonumber\\ 
{\left[X_p^{\downarrow 2}, H_1\right]}_{-}=\sum_{j}\tilde{\epsilon}(pj)
X_j^{\downarrow 2},
\end{eqnarray}
where $\epsilon(pj)$ and $\tilde{\epsilon}(pj)$ are 
non-operator expressions which we calculate using the method 
of work~\cite{did}.
At electron concentration  $n$=1 in a paramagnetic state we have
\begin{eqnarray}
\epsilon(pj)=(1-2d)t_{pj},\nonumber \\
\tilde{\epsilon}(pj)=(1-2d)t_{pj},
\end{eqnarray} 
with $d=\langle X_p^{2}\rangle$ being the concentration of doublons.

Let us take into account
the functions $\langle\langle{\left[X_p^{\downarrow 2}, H'_1\right]}_{-}
|X_{s}^{2\downarrow}\rangle\rangle$ and 
$\langle\langle{\left[X_p^{0 \uparrow}, H'_1\right]}_{-}
|X_{s}^{2\downarrow}\rangle\rangle$ in the mean-field approximation:
\begin{eqnarray}  
\langle\langle{\left[X_p^{\downarrow 2}, H'_1\right]}_{-}
|X_{s}^{2\downarrow}\rangle\rangle&=&
-\sum_{i,i\neq p}t_{ip}
[\langle\langle( X_p^{\downarrow}+X_p^{2}) X_i^{0 \uparrow}
|X_{s}^{2\downarrow}\rangle\rangle
+\langle\langle X_p^{0 2} X_i^{2 \uparrow}
|X_{s}^{2\downarrow}\rangle\rangle
\nonumber\\
&-&\langle\langle X_p^{\downarrow \uparrow} X_i^{0 \downarrow}
|X_{s}^{2\downarrow}\rangle\rangle]
	\simeq -\sum_{i,i\neq p}t_{ip}\langle X_p^{\downarrow}+X_p^{2}\rangle
\langle\langle X_i^{0 \uparrow}|X_{s}^{2\downarrow}\rangle\rangle,
\\
\langle\langle{\left[X_p^{0 \uparrow}, H'_1\right]}_{-}
|X_{s}^{2\downarrow}\rangle\rangle&=&
-\sum_{i,i\neq p}t_{ip}
[\langle\langle( X_p^{0}+X_p^{\uparrow}) X_i^{\downarrow 2}
|X_{s}^{2\downarrow}\rangle\rangle
+\langle\langle X_p^{0 2} X_i^{\downarrow 0}
|X_{s}^{2\downarrow}\rangle\rangle
\nonumber\\
&-&\langle\langle X_p^{\downarrow \uparrow} X_i^{\uparrow 2}
|X_{s}^{2\downarrow}\rangle\rangle]
\simeq-\sum_{i,i\neq p}t_{ip}\langle X_p^{0}+X_p^{\uparrow}\rangle
\langle\langle X_i^{\downarrow 2}|X_{s}^{2\downarrow}\rangle\rangle;
\nonumber
\end{eqnarray}
in this way we neglect the processes describing 
the ``inter-band'' hoppings of electrons which are connected with
spin turning over and ``inter-band'' hoppings with 
creation or annihilation of two electrons on the same site.

So we obtain the closed system of equations 
\begin{eqnarray} \label{syst}
(E-\mu+U)
\langle\langle X_p^{\downarrow 2}|X_{s}^{2\downarrow}\rangle\rangle
-\sum_{i}\tilde{\epsilon}(pi)
\langle\langle X_i^{\downarrow 2}|X_{s}^{2\downarrow}\rangle\rangle
+\langle X_p^{\downarrow}+X_p^{2}\rangle
\sum_{i,i\neq p}t_{ip}
\langle\langle X_i^{\downarrow 2}|X_{s}^{2\downarrow}\rangle\rangle
\nonumber\\
={\langle X_p^{2}+X_p^{\downarrow}\rangle \over 2\pi} \delta_{ps},
\\
(E-\mu)
\langle\langle X_p^{0 \uparrow}|X_{s}^{2\downarrow}\rangle\rangle
-\sum_{i}\epsilon(pi)
\langle\langle X_i^{0 \uparrow}|X_{s}^{2\downarrow}\rangle\rangle
+\langle X_p^{0}+X_p^{\uparrow}\rangle
\sum_{i,i\neq p}t_{ip}
\langle\langle X_i^{\downarrow 2}|X_{s}^{2\downarrow}\rangle\rangle=0.
\nonumber
\end{eqnarray}
After the Fourier transformation we obtain solutions of system of 
Eqs.~(\ref{syst}):
\begin{eqnarray}
\langle\langle X_p^{\downarrow 2}|X_{s}^{2\downarrow}\rangle\rangle_{\bf k}
&=&{\langle X_p^{2}+X_p^{\downarrow}\rangle\over 2\pi}
\left({A^{1}_{\bf k}\over E-E_h({\bf k})}
+{B^{1}_{\bf k}\over E-E_d({\bf k)}}\right),\label{fd}\\
A^{1}_{\bf k}&=&{1\over 2}\left(1-{U-\epsilon({\bf k})
+\tilde{\epsilon}({\bf k})
\over E_d({\bf k})-E_h({\bf k})}
\right), \quad
B^{1}_{\bf k}=1-A^{1}_{\bf k}, \nonumber\\
\langle\langle X_p^{0 \uparrow}|X_{s}^{2\downarrow}\rangle\rangle_{\bf k}&=&
{\langle X_p^{2}+X_p^{\downarrow}\rangle\langle X_p^{0}+X_p^{\uparrow}\rangle
\over 2\pi}\nonumber\\
&&\times {t({\bf k}) \over E_d({\bf k})-E_h({\bf k})}
\left({1\over E-E_h({\bf k})}
-{1\over E-E_d({\bf k)}}\right).
\end{eqnarray}
Here $t({\bf k})$ is the hopping intergral in ${\bf k}-$representation and
\begin{eqnarray}
E_{h}({\bf k})
&=&-\mu+{U \over 2}+{\epsilon({\bf k})+\tilde{\epsilon}({\bf k})\over 2}
\nonumber\\
&-&{1\over 2}{\sqrt{[U-\epsilon({\bf k})+\tilde{\epsilon}({\bf k})]^2
+\langle X_p^{0}+X_p^{\uparrow}\rangle\langle X_p^{\downarrow}+X_p^{2}\rangle
(t({\bf k}))^2}},\\
E_{d}({\bf k})
&=&-\mu+{U \over 2}
+{\epsilon({\bf k})+\tilde{\epsilon}({\bf k})\over 2}\nonumber\\
&+&{1\over 2}{\sqrt{[U-\epsilon({\bf k})+\tilde{\epsilon}({\bf k})]^2
+\langle X_p^{0}+X_p^{\uparrow}\rangle\langle X_p^{\downarrow}+X_p^{2}\rangle
(t({\bf k}))^2}}
\end{eqnarray}
are the energies of electron in lower (``hole'') and upper (``doublon'')
subbands, respectively; $\epsilon({\bf k})$ and $\tilde{\epsilon}({\bf k})$
are the Fourier components of $\epsilon(pj)$ and $\tilde{\epsilon}(pj)$.    

Analogous procedure gives for functions 
$\langle\langle X_p^{\downarrow 2}|X_s^{\uparrow 0}\rangle\rangle$ and
$\langle\langle X_p^{0\uparrow}|X_s^{\uparrow 0}\rangle\rangle$
the following expressions:
\begin{eqnarray}
\langle\langle X_p^{\downarrow 2}|X_s^{\uparrow 0}\rangle\rangle_{\bf k}&=&
\langle\langle X_p^{0 \uparrow}|X_{s}^{2\downarrow}\rangle\rangle_{\bf k},
\nonumber\\
\langle\langle X_p^{0\uparrow}|X_s^{\uparrow 0}\rangle\rangle_{\bf k}
&=&{\langle X_p^{0}+X_p^{\uparrow}\rangle\over 2\pi}
\left({A^{2}_{\bf k}\over E-E_h({\bf k})}
-{B^{2}_{\bf k}\over E-E_d({\bf k)}}\right),\\
A^{2}_{\bf k}&=&B^{1}_{\bf k},\quad B^{2}_{\bf k}=A^{1}_{\bf k}.\nonumber
\end{eqnarray}

Finally, in {\bf k}-representation single-particle Green 
function (\ref{Grf}) we obtain
\begin{eqnarray} \label{elfn}
&&\langle\langle a_{p\uparrow}|a^{+}_{s\uparrow}\rangle\rangle_{\bf k}
={1\over 2\pi} \left({A_{\bf k}\over E-E_h({\bf k})}
+{B_{\bf k}\over E-E_d({\bf k})}\right),\\
&&A_{\bf k}={1\over 2} \left(1-
{(C_1-C_2)
(U-\epsilon({\bf k})+\tilde{\epsilon}({\bf k}))
+4t({\bf k})C_1 C_2
\over E_d({\bf k})-E_h({\bf k})} \right), \nonumber\\
&&B_{\bf k}=1-A_{\bf k},\nonumber 
\end{eqnarray}
where $C_1=\langle X_p^{0}+X_p^{\uparrow}\rangle$,
$C_2=\langle X_p^{2}+X_p^{\downarrow}\rangle$. 

 In the important for an investigation of metal-insulator transition case 
$n=1$ in a paramagnetic state ($\langle X_p^{\uparrow}\rangle=
\langle X_p^{\downarrow}\rangle$) 
single-particle Green function~(\ref{elfn}) has the form 
\begin{eqnarray} \label{elf}
&&\langle\langle a_{p\uparrow}|a^{+}_{s\uparrow}\rangle\rangle_{\bf k}
={1\over 2\pi} \left({A_{\bf k}\over E-E_h({\bf k})}
+{B_{\bf k}\over E-E_d({\bf k)}}\right),\\
&&A_{\bf k}={1\over 2} \left( 1-{t({\bf k})\over 
\sqrt{U^2+(t({\bf k}))^2}} \right), \nonumber\\
&&B_{\bf k}=1-A_{\bf k},\nonumber 
\end{eqnarray}
where single-particle energy spectrum is 
\begin{eqnarray}\label{sp}
&&E_{h}({\bf k})=(1-2d)t({\bf k})
-{1\over 2}{\sqrt{U^2+(t({\bf k}))^2}},\nonumber \\
&&E_{d}({\bf k})=(1-2d)t({\bf k})
+{1\over 2}{\sqrt{U^2+(t({\bf k}))^2}}
\end{eqnarray}
(here we took into account that $\mu={U \over 2}$ for $n=1$).

Single-particle Green function (\ref{elf}) and energy spectrum (\ref{sp})
are exact in the band and atomic limits. It is worthwile to note, that
in distinction from the results of  
two-pole approximations of Hubbard~\cite{hubb1} and Ikeda, Larsen, 
Mattuck~\cite{iked} the energy spectrum (\ref{sp}) depends on polar 
states concentration (thus on temperature). In distinction from 
approximations based on ideology of Roth~\cite{roth} (in this connection 
see also Refs.~\cite{harr}-\cite{izyu}) 
the energy spectrum (\ref{sp}) describes metal-insulator transition.
Energy spectrum which describes metal-insulator transition was 
earlier obtained in work~\cite{did}. Expressions (\ref{sp}) differs from
the respective expressions in work~\cite{did} by
presence of term $\sqrt{U^2+t^2({\bf k})}$ instead of 
$\sqrt{U^2+4d^2t^2({\bf k})}$. This leads to
the series of distinctions between results of this work and results of 
work~\cite{did} ($d(U/w)$--dependence, the condition of metal-insulator 
transition, etc); at the same time expression (\ref{sp}) depends on polar 
state concentration similarly to respective expression in work~\cite{did}.

\section{Energy gap and polar states concentration}
The energy gap (difference of energies between bottom of the upper 
and top of the lower Hubbard bands) is given by 
\begin{eqnarray} \label{gap}
\Delta E=E_d(-w)-E_h(w)=-2w(1-2d)+\sqrt{U^2+w^2},
\end{eqnarray}
(where $w=z|t|$ is the halfwidth of uncorrelated electron band,
$z$ is the number of nearest neighbors to a site).
Expression~(\ref{gap}) describes the vanishing of the energy gap in the
spectrum of paramagnetic insulator at critical value $({U \over w})_c$
when the halfbandwidth $w$ increase (under pressure).
Dependence of $\Delta E$ on temperature can lead to the transition 
from metallic to insulator state with increase of temperature
(in this connection note the transition at increasing temperatures 
from the state of paramagnetic 
metal to the paramagnetic insulator state  
in the systems NiS$_{2-x}$Se$_x$, (V$_{1-x}$Cr$_x$)$_2$O$_3$ and 
Y$_{1-x}$Ca$_x$TiO$_3$). 

For the calculation of polar states concentration we use function~(\ref{fd}). 
At $T=0$ and rectangular density of states the concentration of polar states 
is
\begin{eqnarray} \label{d<}
d={1\over 4}+{U\over 8w}\ln\left({1-4d\over 3-4d}\right)
\end{eqnarray}
if $({U \over w})\leq ({U \over w})_c$ and
\begin{eqnarray} \label{d>}
d={1\over 4}+{U\over 8w}\ln\left({\sqrt{1+{U\over w}^2}+1
\over \sqrt{1+{U\over w}^2}-1}\right)
\end{eqnarray}
if $({U \over w})>({U \over w})_c$. At $T=0$ we have $({U \over w})_c=1.672$.

The dependence $d({U \over w})$ given by Eqs.(\ref{d<})-(\ref{d>}) is plotted
on Fig.~1. One can see that in the point $({U \over w})_c$ the slope 
of $d({U \over w})-$dependence changes; the concentration of doublons
vanishes at \mbox{ ${U \over w}\to\infty$}. Our result for $d({U \over w})$ 
in region of MIT is in good agreement with result of papers~\cite{geo,kotl} 
obtained in the limit of infinite dimensions (Fig.~2). The parameter $U$ is 
normalized by averaged band  energy in absence of correlation $\varepsilon_0$.

In Fig.~3 the dependences of polar states concentration on parameter
${U \over w}$ at different temperatures are presented.
Note the important difference (see Fig.4) of
the dependence of $d$ on temperature from result of papers~\cite{geo,kotl}: 
we found that at any temperature polar states concentration increases 
monotonically with increasing temperature at the fixed value of 
${U \over w}$ when respective dependence in~\cite{geo,kotl}
has a minimum.

The dependence of ${\Delta E\over U}$ on parameter ${U \over w}$ 
at zero temperature is plotted in Fig.~5. It is important to note that in 
the point of gap disappearence $d \neq 0$ in contrast to the previously
obtained result~\cite{did}.
At increasing ${U \over w}$ the energy gap width increases (the negative 
values of $\Delta E$ correspond to the overlapping of the subbands).
For comparison on Fig.5 results of approximation ``Hubbard-I''~\cite{hubb1} 
is also plotted. 
In the point of energy gap vanishing $({U \over w})_c=1.672$ what is very 
close to result of ``Hubbard-III'' approximation~\cite{hubb3}. 

At increase of temperature in metallic state the overlapping of subbands
decreases and temperature induced transition from metallic to insulating
state can occur at some values of parameter ${U \over w}$ (Fig.~6).
The obtained results allows us to draw the $(w/U, T)$ phase diagram of the 
model~(Fig.~7). This phase diagram can explain the experimentally observed 
transitions from metallic to insulating state with increase of temperature
and from insulating to metallic state with increase of bandwidth (under 
external pressure) in paramagnetic state.

\section{Ground state energy}

The ground state energy of the model
\begin{eqnarray} 
{E_0\over N}={1\over N} \langle\sum_{ij\sigma}t_{ij}
a^{+}_{i\sigma}a_{j\sigma}\rangle +Ud,
\end{eqnarray}
calculated using single particle Green function~(\ref{elf}) and
expressions (\ref{d<})-(\ref{d>}) for the concentration of polar states
has the form:
\begin{eqnarray} \label{E0<}
{E_0\over N}=-{w\over 2}+{U\over 4}(1+3d)-{U^2\over 2w}
{(1-4d)\over 4(1-2d)^2-1}
\end{eqnarray}
if $({U \over w})\leq ({U \over w})_c$ and
\begin{eqnarray} \label{E0>}
{E_0\over N}=-{1\over 2}\sqrt{U^2+w^2}+2U({1\over 4}-d)
\end{eqnarray}
if $({U \over w})>({U \over w})_c$.
In Fig.~8 the dependence of the ground state energy on parameter ${U \over w}$ 
given by Eqs.(\ref{E0<})-(\ref{E0>}) is compared with the exact result, 
found in one-dimensional case~\cite{lieb}. The upper and lower bounds on 
ground state energy in one-dimensional case found in paper~\cite{lang} 
are also shown. Our result for the ground state energy in metallic state 
lies slightly lower than exact one and in insulator state fits the exact 
ground state energy very well.

In Fig.~9 our plot of the ground state energy is compared with the best upper 
and lower bounds on ground state energy in infinite-dimensional 
case~\cite{jan}.
In Fig.~10 we have the comparison with bounds on ground state energy for 
three-dimensional simple cubic lattice obtained in paper~\cite{lang}.
In Figs.~8-10 the ground state energy per electron is normalized by
averaged band  energy in absence of correlation $\varepsilon_0$;
in considered case and rectangular density of states 
$\varepsilon_0=-{w \over 2}$.
Figs.~8-10 show that our result present a good approximation for the
ground state energy of the system. In Fig.~11 we plot our result for the 
kinetic part of ground state energy. This plot describes the same behavior 
of kinetic energy of electrons with change of correlation strength in 
paramagnetic state as respective result of work~\cite{kotl}:
in metallic state absolute value of kinetic energy decrease rapidly due to 
rapid  decrease of doublon (hole) concentration.  In insulating state 
absolute value of kinetic energy decrease slowly what in the approximation 
of effective Hamiltonian (obtained for the case ${t_{ij}\over U} \ll 1$)
is equivalent to the interaction of local magnetic moments.
 
\section{Conclusions} 
In this paper we have studied electron correlations in narrow energy bands
using recently proposed approximation~\cite{dcm}. 
We assume that state of the narrow-band system is paramagnetic insulator
or paramagnetic metal.  The single-particle Green function and 
energy spectrum dependent on model parameters and on polar states 
concentration (thus on temperature) have been found
in paramagnetic state at half-filling ($n=1$). The obtained expression 
for energy gap allows to  describe MIT at changes of bandwidth (under external 
pressure) or temperature. The comparison of calculated ground state energy 
with results of other approximations and the exact result found in 
one-dimensional case shows that the used method is
a good approximation for the model under consideration. The obtained phase 
diagram of the model can explain the transitions from 
paramagnetic metal state to paramagnetic insulator state at increase of
temperature and the paramagnetic insulator -  paramagnetic metal transitions 
under external pressure observed in the systems NiS$_{2-x}$Se$_x$, 
(V$_{1-x}$Cr$_x$)$_2$O$_3$ and Y$_{1-x}$Ca$_x$TiO$_3$.

It is worthwhile to note that approximation used in this paper can be 
generalized to describe effects of antiferromagnetic ordering. 
Such a generalization will be considered in subsequent paper.

\newpage

\centerline{\bf Figure captions}

Fig.1
The dependence of doublon concentration $d$ on $U/w$ 
at zero temperature.

Fig.2
The comparison of $d(U/w)$ dependences:
solid line -  our result, dashed line - iterative-perturbative 
theory~[9,23], circles - QMC method~[23].

Fig.3
The dependences of doublon concentration $d$ on $U/w$
at different temperatures: lower curve corresponds to $kT/w=0.16$,
middle curve corresponds to $kT/w=0.08$, upper curve corresponds 
to $kT/w=0$.

Fig.4
The dependences of doublon concentration $d$ on temperature 
at different $U/w$: values of $U/w$ from down to up are 3, 2, 1.5, 1, 0.5, 0

Fig.5
The dependences of energy gap width on $U/w$: 
``Hubbard-I'' approximation (upper curve), our result (middle curve),
approximation~[13] (lower curve).

Fig.6
The dependences of energy gap width on temperature 
at different $U/w$: values of $U/w$ from down to up are 0.5, 1.2, 1.5

Fig.7
The obtained ($kT, w/U$)-phase diagram of the model.

Fig.8
The comparison of ground state energies in one-dimensional case:
dashed curves correspond to upper and lower bounds given by 
Langer and Mattis~[25], 
upper solid curve corresponds to exact ground state (Lieb and Wu~[24])
lower solid curve corresponds to result of this paper.

Fig.9
The ground state energy found in this paper (upper curve),
best upper (middle curve) and lower (lower curve) bounds on 
ground state energy in infinite-dimensional case.

Fig.10
The upper (upper curve) and lower (lower curve) bounds on 
ground state energy in three-dimensional case~[25] and 
the ground state energy found in this paper (middle curve).

Fig.11
The kinetic part  of ground state energy as a function of $U/w$.


\begin{thebibliography}{99}
\bibitem{mott} N.~F.~Mott, Metal-Insulator Transitions, Taylor and Francis,
London 1990.
\bibitem{yao} X.~Yao et al, Phys. Rev. B, {\bf 54}, 17469 (1996); 
Phys. Rev. B {\bf 56}, 7129 (1997).
\bibitem{spa} J.~M.~Honig, J.~Spa\l ek, Chem. Mater., {\bf 10}, 2910 (1998). 
\bibitem{edra} P.~P.~Edvards, C.~N.~R.~Rao, Metal-Insulator Transitions 
Revisited, Taylor and Francis, London 1995.
\bibitem{tag} Y.~Taguchi et al, Phys. Rev. B, {\bf 48}, 511 (1993).
\bibitem{tok} A.~Fujimori, Y.~Tokura, Spectroscopy of Mott Insulators and 
Correlated Metals, Springer-Verlag, Berlin Heidelberg 1995.
\bibitem{hubb1} J.~Hubbard, Proc. Roy. Soc. A{\bf 276}, 238 (1963). 
\bibitem{iz} Yu.~A.~Izyumov, Uspekhi Fizicheskikh Nauk {\bf 165}, 403 (1995) 
(in Russian).
\bibitem{geo} A.~Georges, G.~Kotliar, W.~Krauth, and M.~Rozenberg, Rev. Mod. 
Phys. {\bf 68}, 13 (1996).
\bibitem{gebh} F.~Gebhard, The Mott Metal-Insulator Transition -- Models 
and Methods, Springer, Berlin 1997.
\bibitem{dcm} L.~Didukh, cond-mat/0002334.
\bibitem{hubb} J.~Hubbard, Proc. Roy. Soc. A{\bf 285}, 542 (1965).
\bibitem{did} L.~Didukh, phys. stat. sol.~(b) {\bf206}, R5 (1998).
\bibitem{iked} M.~A.~Ikeda, U.~Larsen, R.~D.~Mattuck, Phys. Lett. {\bf A39},
55 (1972).
\bibitem{roth} L.~Roth, Phys. Rev. {\bf 184}, 451 (1969).
\bibitem{harr} A.~B.~Harris and R.~V.~Lange, Phys. Rev. {\bf 157}, 295 (1967).
\bibitem{kawa} A.~Kawabata, Progr. Theor. Phys. {\bf 48}, 1793 (1972).
\bibitem{been} J.~Beenen, D.~M.~Edwards, Phys. Rev.~B {\bf52}, 13636 (1995).
\bibitem{mehl} B.~Mehlig et al, Phys. Rev.~B {\bf52}, 2463 (1995).
\bibitem{oles} A.~M.~Ole\' s, H.~Eskes, Physica~B {\bf206-207}, 685 (1995).
\bibitem{izyu} Yu.~A.~Izyumov and N.~I.~Chashchin, Fiz. Metal. Metaloved.
{\bf84}, 16 (1997) (in Russian).
\bibitem{hubb3} J.~Hubbard, Proc. Roy. Soc. A{\bf 281}, 401 (1964). 
\bibitem{kotl} G.~Kotliar, M.~Rozenberg, in: The Hubbard Model, edited by 
D.~Baeriswyl, Plenum Press, New York and London 1995, p.155.
\bibitem{lieb} E.~Lieb, F.~Wu, Phys. Rev. Lett. {\bf 20}, 1445 (1968). 
\bibitem{lang} W.~D.~Langer, D.~C.~Mattis, Phys. Lett.~A {\bf 36}, 139 (1971). 
\bibitem{jan} V.~Jani\v s, J.~Ma\v sek, D.~Vollhardt, Z. Phys.~B {\bf 91}, 325 (1993).
\end{thebibliography}
\end{document}